\documentclass[useAMS,usenatbib,usegraphicx]{mn2e}
\def\apj #1 {{ApJ,} { #1}, }
\def\apjs #1 {{ApJS,} { #1}, }
\def\apjsupp #1 {{ApJS,} { #1}, }
\def\aj  #1 {{AJ,} { #1}, }
\def\aa  #1 {{A\&A,} { #1}, }
\def\aap  #1 {{A\&A,} { #1}, }
\def\aarev  #1 {{A\&AR,} { #1}, }
\def\aasupp  #1 {{A\&AS,} { #1}, }
\def\araa  #1 {{ARA\&A,} { #1}, }
\def\mnras  #1 {{MNRAS,} { #1}, }
\def\mn  #1 {{MNRAS,} { #1}, }
\def\pasp  #1 {{PASP,} { #1}, }
\def\pasa  #1 {{PASA,} { #1}, }
\def\paspc  #1 {{PASPC,} { #1}, }
\def\pasj  #1 {{PASJ,} { #1}, }
\def\revmex  #1 {{Rev. Mex. Astr. Astrofis.,} { #1}, }
\def\revmexconf  #1 {{Rev. Mex. Astr. Astrofis. Conf. Ser.,} { #1}, }
\def\jaa #1 {{JA\&A,} { #1}, }
\def\annrev #1 {{ARA\&A,} { #1}, }
\def\apss #1 {{Ap\&SS,} { #1}, }
\def\baas #1 {{BAAS,} { #1}, }
\def\nat #1 {{Nature,} { #1}, }
\def\obs #1 {{Observatory,} { #1}, }
\def\memital #1 {{Mem. Soc. Astron. Ital.,} { #1}, }
\def\josa #1 {{J.Opt.Soc.Am.,} { #1}, }
\def\sci #1 {{Science,} { #1}, }
\def\etal {{et al.~}}
\def\lsol{L_\odot}                  % solar luminosity
\def\msol{M_\odot}                  % solar mass           
\def\rsol{R_\odot}                  % solar radius
\def\mic{\,\mu \rm m}                           % micron
\def\1{\'\i}
\def\mic{\,\mu \rm m}
\title[]{An infrared study of the high-mass, multi-stage star-forming region IRAS~12272-6240\thanks{Based on observations 
made with the 6.5m Baade telescope of Las Campanas Observatory, Chile. {\sl Herschel} is an ESA space observatory with 
science instruments provided by European-led Principal Investigator consortia and with important participation from NASA. }}

\author[M. Tapia et al.]
{Mauricio Tapia$^{1}$\thanks{E-mail: mt@astro.unam.mx}, 
Paolo Persi$^{2}$\thanks{E-mail: paolo.persi@inaf.it},
Miguel Roth$^{3,4}$, Davide Elia$^{2}$ \\
$^{1}${Instituto de Astronom{\'\i}a, Universidad Nacional Aut\'onoma de M\'exico, Ensenada, B. C., CP 22830, Mexico}\\
$^{2}${INAF-Istituto Astrofisica e Planetologia Spaziale, Via Fosso del Cavaliere 100, 00133 Roma, Italy}\\
$^{3}${Las Campanas Observatory, Carnegie Institution of Washington, La Serena, Chile}\\
$^{4}${Giant Magellan Telescope Organization, Pasadena, USA}}

\pagerange{\pageref{firstpage}--\pageref{lastpage}} \pubyear{2019}

\begin{document}

\date{Accepted 2020  . Received  2020 ?; in original form 2019 }

\label{firstpage}

\maketitle
\begin{abstract}

IRAS 12272-6240 is a complex star forming region with a compact massive dense clump and several associated masers, located at a well-determined 
distance of $d=9.3$ kpc from the Sun. For this study,  we obtained sub-arcsec broad- and narrow-band near-IR imaging and low-resolution 
spectroscopy with the Baade/Magellan telescope and its camera PANIC. Mosaics of size $2 \times 2$ square arcmin in the $JHK_s$ 
bands and with narrow-band filters centred in the 2.12 $\mu$m H$_2$ and 2.17  $\mu$m Br$\gamma$ lines were analysed in combination 
with HI-GAL/{\sl Herschel} and archive IRAC/{\sl Spitzer} and {\sl WISE} observations.  We found that the compact dense clump houses  
two Class~I YSOs that probably form a 21 kAU-wide binary system. Its combined 1 to 1200 $\mu$m SED is consistent with an O9V central 
star with a $10^{-2} \msol$ disc and a $1.3 \times 10^4 \msol$ dust envelope. Its total luminosity is $8.5 \times 10^4 \lsol$. 
A series of shocked H$_2$ emission knots are found in its close vicinity, confirming the presence of outflows. 
IRAS 12272-6240 is at the centre of an embedded cluster with a mean age of 1 Myr and 2.6 pc in size that contains more than 150 stars. 
At its nucleus, we found a more compact and considerably younger sub-cluster containing the YSOs. We also identified and classified 
the O-type central stars of two dusty radio/IR HII regions flanking the protostars. Our results confirm that these elements form a single giant 
young complex where massive star formation processes started some 1 million years ago and is still active.
% 248 words
\end{abstract}


\section{Introduction}

It is widely accepted that understanding the complex processes that result in the formation of massive stars (O, B-stars with masses $\ge 8 \msol$)  
requires, apart from comprehensive physical principles, knowledge of the precise characteristics of the environments that lead to such events. 
These vary considerably from one region to the next and, thus, it is vital to determine observationally the detailed conditions of a wide 
collection of natal stellar regions in our Galaxy.  This paper is a continuation of such efforts, mainly through the use of multi-band 
calibrated imaging from near-infrared (near-IR) to millimetre wavelengths complemented by spectroscopic data (e.g. Tapia \etal 2014, Persi \etal 2016, 
Tapia et al. 2018, Persi \& Tapia  2019). 

\begin{figure*}
\resizebox{18.26cm}{!}{\includegraphics[clip=true]{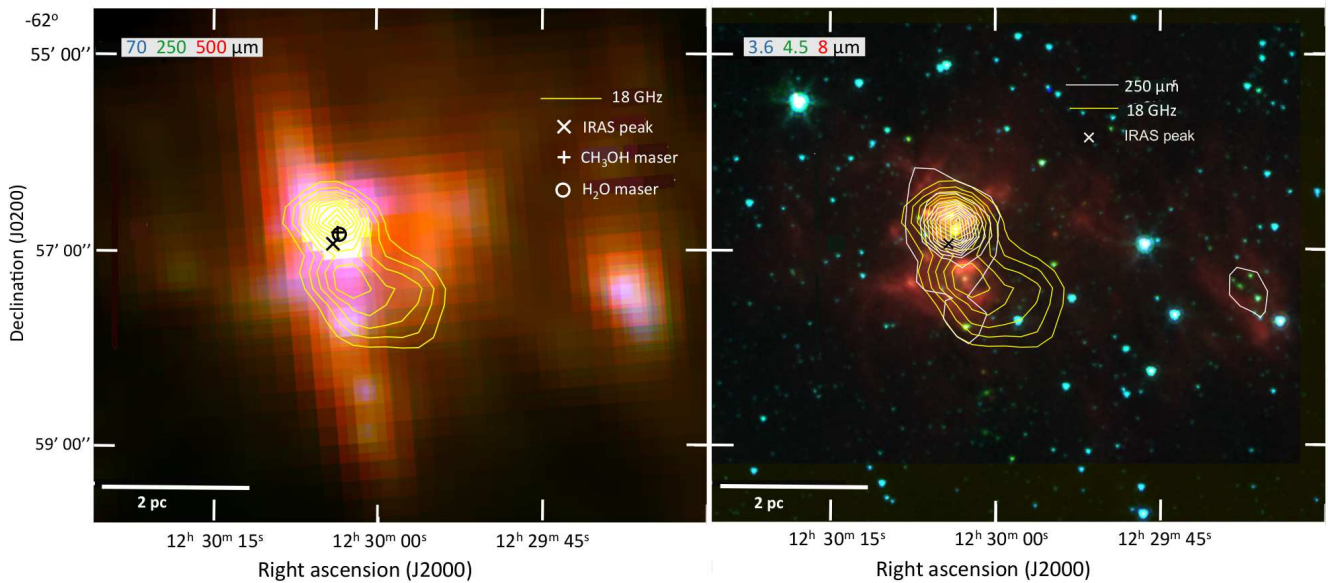}}
\caption{Mid- and far-infrared composite colour images of the region centred on IRAS 12272-6240. The left panel shows the image constructed from the HI-GAL 
{\sl Herschel}  PACS  70~$\mic$ (blue), SPIRE 250~$\mic$ (green) and 500~$\mic$ (red) frames. The positions of the methanol and water masers (Breen 
et al. 2010; Sanchez-Monge \etal 2013), and of the IRAS peak are indicated. The right panel shows the  {\sl Spitzer} IRAC  image constructed from the 3.6~$\mic$ 
(blue), 4.5~$\mic$ (green) and 8~$\mic$ (red) frames. The yellow contours correspond to the radio emission at 18GHz observed by 
Sanchez-Monge \etal (2013) and the white contours correspond to the HI-GAL 250~$\mic$  emission. The scale is for a distance of 9.3 kpc.}
\label{fig1}
\end{figure*}

One of the main differences between high-mass and low-mass stars is that the radiation field of a massive star plays a more important role during 
its whole life, including its formation stages. Theoretically, a massive protostellar embryo heats up and eventually ionizes the gas of 
its surrounding envelope, creating an HII region which develops by expanding within the cloud. Different theories of the high mass star formation 
have been proposed by several authors (see reviews by,  e.g.,  Beuther \etal 2007, Tan \etal 2014).

IRAS12272-6240 is a complex star forming region, classified as a ``high'' source by Palla et al. (1991) from its [25-12] and [60-12] IRAS colours.  
Three dense clumps were detected at 1.2~mm by Beltran \etal (2006), the brightest of which is close to the nominal IRAS position.  6.7~GHz methanol 
and  22~GHz water masers were detected by Pestalozzi \etal (2005), Caswell (2009), Breen \etal (2010), and S\'anchez-Monge et al. (2013), close to the 
position  of an OH maser  (Caswell \etal  1998, Breen et al. 2010). In fact, all of them are associated with the millimetre emission peak dust clump. All these
 observations confirm that IRAS12272-6240 is a high mass young stellar object (YSO; cf. Lumsden \etal 2013).

There are several published kinematic distance estimates to this star formation complex, all based on radial velocities of masers, molecular
and atomic lines. Beltr\'an et al. (2006),  derived a value of $d = 11.2$ kpc. More recent determinations resulted in smaller  distances to  IRAS12272-6240. 
The Red MSX Source survey (Lumsden et al. 2013) lists a value of  $d = 8.9$ kpc. The APEX Telescope Large Area Survey of the inner Galactic plane 
at 870 $\mu$m (ATLASGAL) kinematic distances catalogue (Wienen et al. 2015)  lists $d = 9.71 \pm 0.73$ kpc. More recently, Whitaker et al. (2017), 
also using multiple molecular-line radial velocity data from the Millimetre Astronomy Legacy Team 90 GHz Survey (MALT90, Rathborne et al. 2017), 
determined an improved value of $d = 9.39 \pm 0.62$ kpc for this region. Throughout this work, we will assume a weighted mean value of $d = 9.3$ kpc.
Given its Galactic coordinates $l=300.50,~b=0.18$, this region is located at the far end of the Carina-Sagittarius arm, at a Galactocentric 
radius of $R_{\rm gal} = 9.0$ kpc.

In order to perform a detailed analysis of this young region, we have obtained new sub-arcsec resolution near-IR broad-band and narrow-band images 
centred on the IRAS source. We also gathered near-IR spectra of two bright sources. These observations were analysed together with {\sl Herschel} 
images obtained from the {\sl Herschel} Infrared GALactic Plane survey (Hi-GAL, Molinari \etal  2010) and with archived IRAC/{\sl Spitzer} images.

All observational material and reduction methods are described in Section 2, while in Section 3 we present the results and discuss the main 
derived properties of this high mass star forming region. Finally  Section 4 lists our conclusions.

\section[]{Observations}

\subsection{Hi-GAL images and photometry}
 
\begin{table*}
\caption[] { Observed {\sl Herschel} flux densities  of IRAS 12272-6240 and derived parameters}
\begin{tabular}{ccccccccc}
\hline
 $\alpha (2000)$ & $\delta (2000)$  & F[70]& F[160] & F[250] & F[350] & F[500] & M & T$_d$ \\  
   h\quad m\quad s & $\circ \quad ^{\prime }\quad ''$& Jy & Jy & Jy & Jy & Jy & $M_\odot$ & $K$\\  
 \hline  
     12 30  03.7 & -62  56 49 &   $610.2\pm 22.0$ & $490.0\pm 36.2$  & $340.5 \pm 40.8$& $107.6\pm 30.3$ &$ 51.1\pm 10.8$ & $8850 \pm 80$ &$19.8 \pm 0.7$  \\
\hline 
\end{tabular}
\end{table*}
% Table 1

Far-infrared images at 70, 160, 250, 350 and 500~$\mu$m of the region centred at IRAS 12272-6240 have been obtained as part of the {\sl Herschel} 
Hi-GAL survey. The images were reduced using the HI-GAL standard pipeline (Traficante \etal 2011). Source extraction and photometry was done using 
the Curvature Threshold Extractor package (CuTEx, Molinari \etal 2011) independently at each band. This package performs a fit of a 2-D Gaussian to the
source profile, and the flux uncertainties are essentially related to the quality of such a fit. 

Figure 1 (left panel) illustrates the {\sl Herschel} three-colour image centred around the IRAS position. This is composed of the 70~$\mu$m (blue), 
250~$\mu$m (green) and 500~$\mu$m (red) frames.  Table~1 lists the position and the measured flux densities, $F_{\nu}$, of IRAS 12272-6240. From these, 
the far-IR spectral energy distribution (SED) of IRAS 12272-6240, illustrated in Fig.~2, was constructed. We then derived the dust mass and 
temperature of this dense clump by fitting a single-temperature modified black body to the SED, following Giannini \etal (2012) and Elia \etal (2013).  
The resulting parameters, for a distance  $d = 9.3$ kpc, are also reported in Table~1.
         
\begin{figure}
\resizebox{8.2cm}{!}{\includegraphics[clip=true]{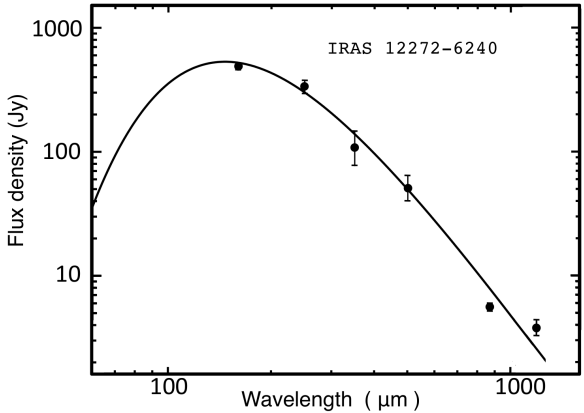}}
\caption{ Far-IR  and millimetre spectral energy distribution (SED) of IRAS 12272-6240. The continuous line is the best-fit modified black body  
described in Section 2.1. The 870~$\mu$m APEX/ LABOCA flux is from Csengeri et al. (2014) and fluxes in the wavelength range 
160 to 500 ~$\mu$m  are  from {\sl Herschel}/HI-GAL. The $70 ~\mu$m flux was not considered for the fit. For completeness, the 1.2mm flux 
from Beltr\'an et al. (2006) is also drawn.}
\label{fig2}
\end{figure}

\subsection{{\sl Spitzer} GLIMPSE and {\sl WISE} archive images.}

Flux-calibrated images of the IRAS 12272-6240 region from the {\sl Spitzer} Galactic Legacy Infrared Mid plane Survey Extraordinaire (GLIMPSE; 
Benjamin \etal 2003; Churchwell \etal 2009) key program survey taken at 3.6, 4.5, 5.8 and 8 $\mu$m with IRAC (Fazio \etal 2004) on board 
the {\sl Spitzer} Space Telescope (Werner \etal 2004) were retrieved from the public image archive at National Aeronautics and Space 
Administration/Infrared Processing and Analysis Center (NASA/IPAC) Infrared Science Archive (IRSA). 
The colour-composite image, made up of the IRAC 3.6, 4.5 and 8 $\mu$m frames of the surveyed area, is presented in Fig.~1 (right panel), 
and scaled up, in Fig.~3 (right panel). The flux at 8 $\mu$m is dominated by polycyclic aromatic hydrocarbon (PAH) emission. At the position of 
the {\sl Herschel} dense core, a bright mid-IR source is reported in the GLIMPSE catalogue (G300.5039-001759). Mercer \etal (2005) reported 
an overdensity of GLIMPSE sources in this area (number 33 in their Table~1), suggesting the presence of a young star cluster of size (diameter)
$\sim 48\arcsec$  and containing around 34 young mid-IR sources. 

We performed $4\arcsec$-aperture photometry for all the sources which were bright enough to get reliable photometry, that is, with intrinsic errors less 
that 15\%. The photometric zero points for the four IRAC channels were derived from the point sources in this field included in the GLIMPSE 
catalogue. These are indicated in Table~2 . The positions and photometry, including the corresponding $JHK_s$ magnitudes (see Section 2.3), 
are also shown in Table~2, while Fig.~4 displays the corresponding $H-K_s$ versus $K-[3.6]$ and [3.6] - [4.5] versus [4.5] - [5.8] diagrams. 

\begin{figure*}
\resizebox{18.3 cm}{!}{\includegraphics[clip=true]{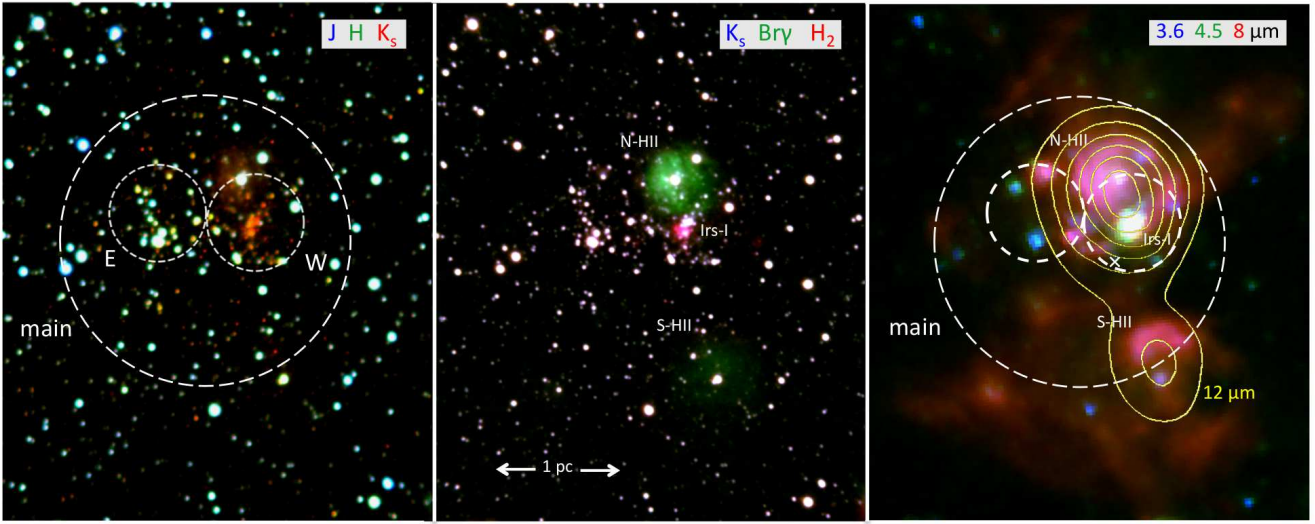}}
\caption{Composite PANIC  $JHK_s$  (left panel) and  $K_s$, Br$\gamma$, and H$_2$ at  2.12 $\mu$m (central panel), and 
{\sl Spitzer}/IRAC 3.6, 4.5, 8 $\mu$m (right panel)  images (blue-green-red, respectively) of the $82\arcsec \times 100\arcsec$ 
area around IRAS12272-6240. The scale is for a distance of 9.3 kpc. The broken-line circles illustrate the centres and sizes of the 
``main'' cluster, of the eastern (E), and of the western (W) subclusters (cf. Table 6). The northern (N) and southern (S)  HII regions, 
and the double protostellar system Irs-1 are indicated. The yellow contours represent the {\sl WISE} 12 $\mu$m emission. 
The cross marks the IRAS peak. The common centre of the images is at 
(J2000)$\alpha  = 12^{\rm h}~30^{\rm m}~04\rlap.{^{\rm s}}6$, $\delta = -62^\circ~56'~55''$. North is to the top and east to the left.}
\label{fig3}
\end{figure*}

To cover the 12 and 22 $\mu$m wavelength region, we also retrieved images of the region from the Wide-field Infrared Survey Explorer mission 
({\sl WISE}; Wright et al. 2010), specifically from the AllWISE Image Atlas archive. The flux densities were retrieved from the AllWISE Source 
Catalog at IRSA. Additionally, fluxes in this wavelength range were taken from the $MSX$ point source catalogue (Egan et al. 2003).
These data are listed in the footnotes of Table 2.

 \subsection{Near-infrared images.}

Our near-infrared images of IRAS 12272-6240 were obtained through standard broad-band $JHK_s$ filters as well as through narrow-band filters 
centred on the lines of H$_2$ ($\lambda$o = 2.122~$\mic$,  $\Delta \lambda = 0.024~\mic$) and Br$\gamma$ ($\lambda$o = 2.165 $\mu$m, 
$\Delta \lambda = 0.022~\mic$). The observations were done on the night of 2009 June 12 using the Perssons Auxiliary Nasmyth Infrared Camera (PANIC) 
attached to the Magellan Baade 6.5 m telescope at Las Campanas Observatory (Chile). PANIC uses a Hawaii 1024 $\times$ 1024 HgCdTe array that 
provides, once mosaiced,  a  $120\arcsec \times 120\arcsec$ field of view with a scale of $0.125\arcsec$/pix (Martini et al. 2004). We obtained 
nine dithered  frames, each of 60, 40 and 20 s effective integration time in $J$, $H$, $K_s$, respectively, and 60 s in the narrow-band filters, by offsetting 
the telescope by 6$\arcsec$ between consecutive exposures. The mean measured full width at half-maximum (FWHM) of the point spread function (PSF) 
in $K_s$  was $0.5\arcsec$. For the narrow-band 2.12 and 2.17 $\mu$m, the measured FWHM was $0.7\arcsec$. 

Figure 3 shows three composite colour near-IR images of the studied region. The left panel illustrates the image obtained by combining the J (blue), H (green) 
and $K_s$ (red) individual frames. The colour image presented in the middle panel was constructed from the $K_s$ broad-band frame (blue) and the narrow-band 
images centred in Br$\gamma$ (green) and the H$_2$ (red) lines at 2.17 and 2.12 $\mu$m, respectively. The right panel shows, at the same scale, the IRAC
3.6, 4.5, 8 $\mu$m composite image with the {\sl WISE} 12 $\mu$m contours for comparison.

PSF-fitting $JHK_s$ photometry was performed using the DAOPHOT Stellar Photometry Package (Stetson, 1987) within the Image Reduction and Analysis 
Facility (IRAF) environment. The photometry was calibrated using several standard stars each night from the extended list of faint standards of 
Persson et al. (1998) for use with the Magellan telescopes. 

A total of 1150 sources were measured in the $H$ and $K_s$ filters in the surveyed area of 4 square arcmin. For the analyses,
we limited the sample to those with intrinsic photometric uncertainties of less than 12\% in $H$ and less than 10\% in $K_s$, while for the
$J$-band analysis, we considered only those measurements with uncertainties less than 12\%, or $J<19.7$. This process left our study sample
with 740 and 1053 sources with reliable $JHK_s$, and $HK_s$ photometry, respectively. From source counts, we estimated a $K_s$-band 
completeness limit (90\%)  to be 17.8, though a significant number of sources had measurements with $17.8 < K_s < 18.5$. For $H$, 
this was 19.2. The full tables listing all  the individual $JHK_s$ photometry (Tables A1 and A2) are only available electronically through 
CDS.\footnote{http://cdsarc.u-strasbg.fr/viz-bin/qcat?J/MNRAS/}. For those sources with mid-IR counterparts on the IRAC images, the  $JHK_s$ 
photometry is listed in Table~2;  in the case of those bright enough to be saturated on any of our PANIC images, the 2MASS Point Source
Catalogue photometry in the three bands is quoted.

\begin{figure*}
\center
\resizebox{17.6 cm}{!}{\includegraphics[clip=true]{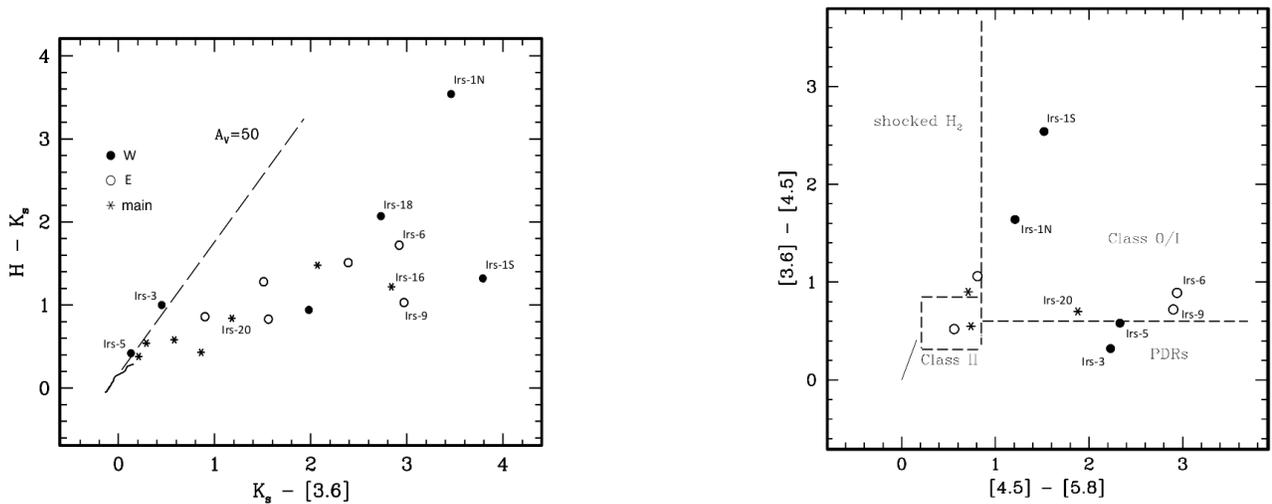}}
\caption{The left panel shows the $H - K_s$ versus $K_s - [3.6]$ plot for all unresolved sources measured in these bands. 
The solid line close to the origin represents the locus of the  main sequence (Koornneef 1983) and the dashed line represents the reddening vector 
(Tapia 1981) of length  $A_V=50$  (Rieke \& Lebofsky 1985). The right panel shows the [3.6] - [4.5] versus [4.5] - [5.8] diagram of the sources with IRAC photometry.  
The broken lines delineate the areas occupied by Class~0/I and Class~II objects. Also indicated with labels are the approximate zones occupied by 
shocked H$_2$- and  PAHs-dominated emission regions  (Ybarra et al. 2013). Filled circles are for the W cluster, the open circles are for the E cluster and the asterisks, 
for the rest of the main cluster.  The labels refer to the identifications in Table~2, which lists the individual photometric values.}
\label{fig4}
\end{figure*}

Analysing in detail the images presented in Fig. 3, we notice the following features:

1. In the near-IR images (left and centre panels) it is obvious that there is an overdensity of sources around the centre of the observed field. Conspicuously,
two star-density peaks are apparent; one dominated by very red (i.e. $K$-flux dominated) sources (hereafter named W cluster), 
and a second one (named E) dominated by brighter, bluer sources, some $18\arcsec$ to the east of the former. Precise star-counts down to our
$K$-band completeness limit (see Section 3.3) indicate that the cluster (hereafter named ``main'') reaches a radius of $28\arcsec$. 
We interpret these features as evidence of the presence of one large embedded cluster with two smaller sub-clusters.  
Their properties will be discussed in detail in Section 3.3.

2. At the centre of the W cluster, within the dense core close to the water, OH and methanol masers, we register the presence of a bright, extended H$_2$ 
emission feature (red in the middle panel of Fig.~3 and in greater detail in Fig.~5), that appears associated with the massive double Class~I YSO system Irs-1S and Irs-1N.
In its vicinity, we also found a number of other faint molecular hydrogen emission knots.  The observed characteristics of this star-formation site 
will be discussed in detail in Section 3.1.

3. Two round HII regions (marked N and S) that emit strongly in the Br$\gamma$ line at 2.17 (green in the middle panel of Fig.~3) are evident. Both have 
diameters of around $13\arcsec$ and have distinct polycyclic aromatic hydrocarborn (PAH)-emission counterparts at 8 $\mu$m (red in the right panel of Fig.~3).
Each HII region has a relatively near-IR-bright star in the centre providing the ionising energy. The details will be presented in Section 3.2. 

\begin{figure*}
\resizebox{15.7 cm}{!}{\includegraphics[clip=true]{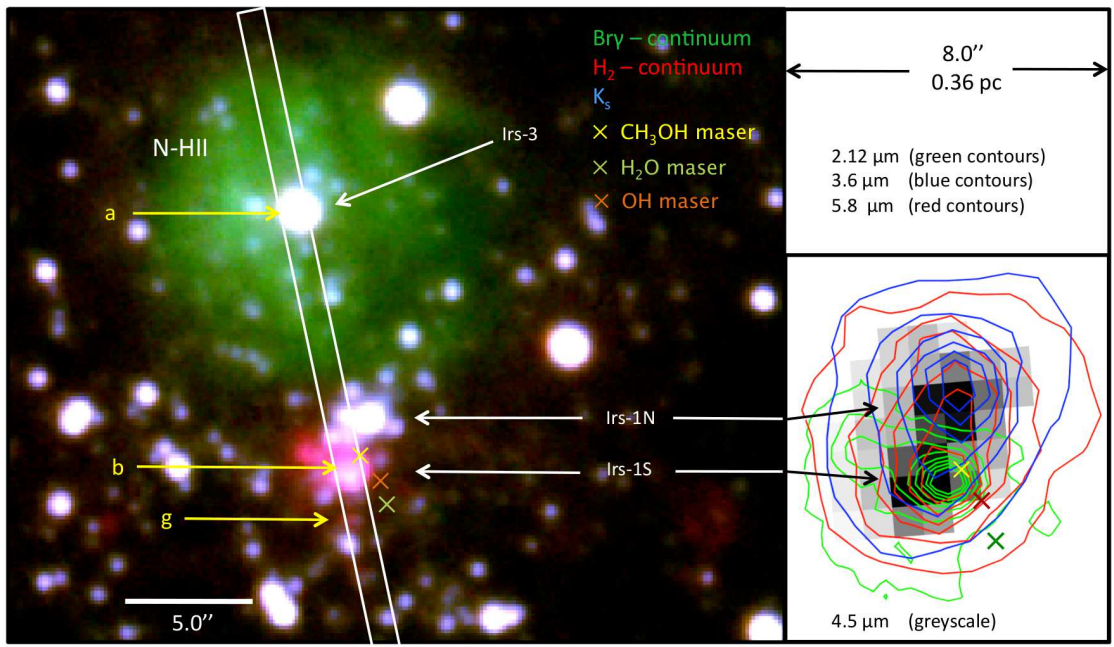}}
\caption{Close-up images of the central zone of IRAS12272-6240. The colour image is coded $K_s$ in blue, Br$\gamma$ in green and H$_2$ in red.  
The  spectrograph's slit position and width are illustrated by a white rectangle. Labels ``a'' and ``b''  mark the locations on the slit where the spectra 
were extracted.  The right panel shows, magnified, a section of the IRAC images centred on Irs~1 at 4.5 $\mu$m (greyscale), 5.8 $\mu$m (red contours), 
and 3.6 $\mu$m (blue contours) while the green contours represent the H$_2$ emission at  2.12 $\mu$m.  The crosses mark the precise positions of the 
methanol (CH$_3$OH), water (H$_2$O), and hydroxide (OH) masers (Breen et al. 2010, S\'anchez-Monge et al. 2013, Caswell 2009). 
The indicated scale is for a distance of 9.3 kpc.}
\label{fig5}
\end{figure*}

\setlength{\tabcolsep}{4pt}

\begin{table*}
\caption[] {Coordinates and photometry of the sources with IRAC photometry.}
%\begin{tabular}{lcccccccccl}
\begin{tabular}{llllllllll}
\hline
Ident. & R.A.~(J2000) & ~Dec. & ~~[3.6] & ~~[4.5] & ~~[5.8] & ~~[8] & ~~~$J$ & ~~~$H$ & ~~~$K_s$  \\  
  Irs & h\quad m\quad s & $^\circ \quad ^{\prime }\quad ''$ & ~(mag) & ~(mag) & ~(mag) & (mag) & ~(mag) & ~(mag) & ~(mag)   \\  
 \hline  
%       ar           dec       ch1    ch2     ch3   ch4     J      H      K      GLIMPSE ID         
1N & 12 30 03.53 & -62 56 47.2 & $10.61 \pm .14$ &  $~~8.97 \pm .09$ &  $~~7.76 \pm .12$ & $6.38 \pm .09$ & $21.21 \pm .11$ & $17.61 \pm .06$ & $14.07 \pm .05$ \\  
1S & 12 30 03.64 & -62 56 49.4 & $11.48 \pm .17$ &  $~~8.94 \pm .09$ &  $~~7.42 \pm .10$ &  $5.95 \pm .09$ & $19.53 \pm .08$ & $16.59 \pm .05$ & $15.27 \pm .05$ \\
3$^\dagger$ & 12 30 03.95 & -62 56 39.1& $11.95 \pm .20$ & $11.63 \pm .14$  &  $~~9.40 \pm .17$ &  $7.19 \pm .07$ & $14.72 \pm .01$* & $13.40 \pm .01$* & $12.40 \pm .01$*  \\
4 & 12 30 03.34 & -62 56 35.0 & $12.17 \pm .19$ & $11.27 \pm .15$ & $10.56 \pm .22$ &   ~~~~-   & $14.49 \pm .01$* & $13.33 \pm .01$* & $12.75 \pm .01$*   \\      
5 & 12 30 02.42 & -62 56 44.5 & $12.61 \pm .20$ & $12.03 \pm .21 $ & $~~9.70 \pm .15$ &   ~~~~-   & $14.46 \pm .01$* & $13.31 \pm .01$* & $12.73 \pm .01$*   \\     
6 & 12 30 05.35 & -62 56 50.5 & $12.56 \pm .20^+$  & $11.67 \pm .17^+$ & $~~8.73 \pm .17^+$ &   ~~~~-   & $18.28 \pm .05$ & $17.20 \pm .10$ & $15.48 \pm .07$   \\    
7 & 12 30 05.23 & -62 56 46.8 & $13.17 \pm .24$ & $11.55 \pm .16$ &   ~~~~-   &   ~~~~-   & $17.04 \pm .01$ & $15.56 \pm .02$ & $14.73 \pm .01$ \\      
8 & 12 30 06.31 & -62 56 51.3 & $11.31 \pm .19^+$ & $10.79 \pm .14^+$ & $10.23 \pm .09^+$ & $ 6.95 \pm .11^+$ & $14.46 \pm .01$* & $13.07 \pm .01$* & $12.20 \pm .01$*   \\
9 & 12 30 06.02 & -62 56 38.3 & $12.03 \pm .20$ & $11.31 \pm .16$ & $~~8.41 \pm .09$ & $6.71 \pm .08$ & $17.86 \pm .02$ & $16.03 \pm .02$ & $15.00 \pm .03$ \\      
10 & 12 30 06.92 & -62 56 40.9 & $11.98 \pm .15^+$ & $10.92 \pm .11^+$ & $10.11 \pm .14^+$ &   ~~~~-   & $17.75 \pm .02$ & $15.88 \pm .02$ & $14.37 \pm .02$   \\    
11 & 12 30 05.34 & -62 56 34.3 & $13.81 \pm .25$ & $11.96 \pm .18$ &   ~~~~-   &   ~~~~-   & $18.80 \pm .11$ & $16.60 \pm .02$ & $15.32 \pm .03$ \\      
12 & 12 30 08.56 & -62 56 54.0 & $12.28 \pm .17^+$ & $12.37 \pm .21^+$ &   ~~~~-   &   ~~~~-   & $14.21 \pm .01$* & $13.11 \pm .01$* & $12.57 \pm .01$*   \\
13 & 12 30 02.87 & -62 56 55.1 & $12.35 \pm .13^+$ & $11.80 \pm .16^+$ & $11.06 \pm .20^+$ &  ~~~~-   & $17.89 \pm .02$ & $15.90 \pm .02$ & $14.42 \pm .02$   \\     
14 & 12 30 01.31 & -62 56 55.7 & $13.05 \pm .09^+$ & $13.15 \pm .12^+$ &   ~~~~-   &   ~~~~-   & $14.73 \pm .01$ & $13.64 \pm .01$ & $13.26 \pm .02$   \\    
15 & 12 30 03.38 & -62 57 00.3 & $12.94 \pm .19$ & $11.50 \pm .12$ &   ~~~~-   &   ~~~~-  & $15.35 \pm .02$ & $14.23 \pm .02$ & $13.80 \pm .02$ \\  
16 & 12 30 06.33 & -62 57 03.8 & $12.54 \pm .17$ & $12.13 \pm .14$ &   ~~~~-   &   ~~~~-  & $18.64 \pm .03$ & $16.60 \pm .02$ & $15.38 \pm .02$ \\        
18 & 12 30 03.34 & -62 56 44.0 & $12.90 \pm .19$ & $11.17 \pm .08$ &   ~~~~-   &   ~~~~-  & $20.28 \pm .06$ & $17.70 \pm .03$ & $15.63 \pm .03$ \\        
19 & 12 30 04.08 & -62 56 54.7 & $12.20 \pm .14$ & $12.68 \pm .21$ &   ~~~~-  &   ~~~~-  & $17.02 \pm .02$ & $15.12 \pm .02$ & $14.18 \pm .02$ \\        
20$^\ddagger$ & 12 30 02.79& -62 57 18.0& $12.06 \pm .11$ & $11.36 \pm .20$ & $ ~~9.48 \pm .11$ & $ 7.50 \pm .02$ & $15.69 \pm .02$ & $14.08 \pm .02$ & $13.24 \pm .02$  \\
\hline 
\multicolumn{10}{l} {NOTES: ~$\dagger$ Central star of Northern HII region;~~~~~~~~~~~~~~~~~~ $\ddagger$ Central star of Southern HII region; } \\ 
\multicolumn{10}{l} {~~~~~~~~~~~~~* $JHK_s$ Photometry quoted from 2MASS; ~~~~~~~~~~  $^+$ Photometry quoted from GLIMPSE. } \\
\multicolumn{10}{l} {~~~~~~~~~~~ $WISE$ fluxes Irs1N+Irs1S: $4.31 \pm 0.09$ Jy at 12$\mu$m ($7''$ beam); $60.33\pm 1.90$ Jy at  22$\mu$m ($12''$ beam)} \\
\multicolumn{10}{l} {~~~~~~~~~~~ $MSX$~ fluxes Irs1N+Irs1S:  $4.45 \pm 0.02$ Jy at 12$\mu$m; $6.58 \pm 0.40$ Jy at 15$\mu$m ($9''$ beam); $34.1\pm 2.0$ Jy at  21$\mu$m ($15''$ beam)} \\
\multicolumn{10}{l} {~~~~~~~~~~~ $WISE$ fluxes Irs3:~~~~~~~~~~~ $1.59 \pm 0.07$ Jy at 12$\mu$m ($7''$ beam); $27.93\pm 0.60$ Jy at  22$\mu$m ($12''$ beam)} \\
\end{tabular}
\end{table*}

\setlength{\tabcolsep}{9pt}

 \subsection{Near-infrared spectroscopy.}
 
Low-resolution long-slit spectroscopy was performed  with the Folded-port Infrared Echellette Spectrograph (FIRE) mounted on the 6.5-m 
Magellan/Baade telescope at  Las Campanas Observatory in its high throughput prism mode. This configuration provides simultaneous spectra 
from 0.82 to 2.51 $\mu$m with spectral resolutions R$_J$ = 500, R$_H$ = 450 and R$_K$ = 300 in the $J$, $H$, $K_s$ atmospheric windows. 
The instrument is described in detail by Simcoe \etal (2013). The fixed effective slit length is 44$\arcsec$ and its width was set to 0.8$\arcsec$, 
which is larger than the typical seeing encountered during our run. The spatial scale is 0.15$\arcsec$/pixel. 
The spectra were reduced using the FIREHOSE package, which is based on the MASE and SpeX reduction tools (Vacca,  Cushing \& 
Rayner 2003, Cushing, Vacca \& Rayner 2004, Bochanski et al. 2009).

We show in Figure 5 the set position of the spectrograph slit in order to cover several of the $K$-band brightest features, in particular the central star 
of the northern HII region Irs-3 (labelled a), and the protostar Irs-1S (labelled b). All spectra reduction procedures for extraction, wavelength  
calibration, sky subtraction, averaging of individual spectra, corrections for telluric absorptions and flux calibration, were done using the 
standard FIRE pipeline reduction described in http://web.mit.edu/rsimcoe/~www/FIRE/obdata.htm. Final filtering for spurious spikes, 
and flux measurements (with Gaussian fitting) was performed with IRAF ONEDSPEC package.

\begin{figure*}
%\hbox{
\resizebox{18.2cm}{!}{\includegraphics[clip=true]{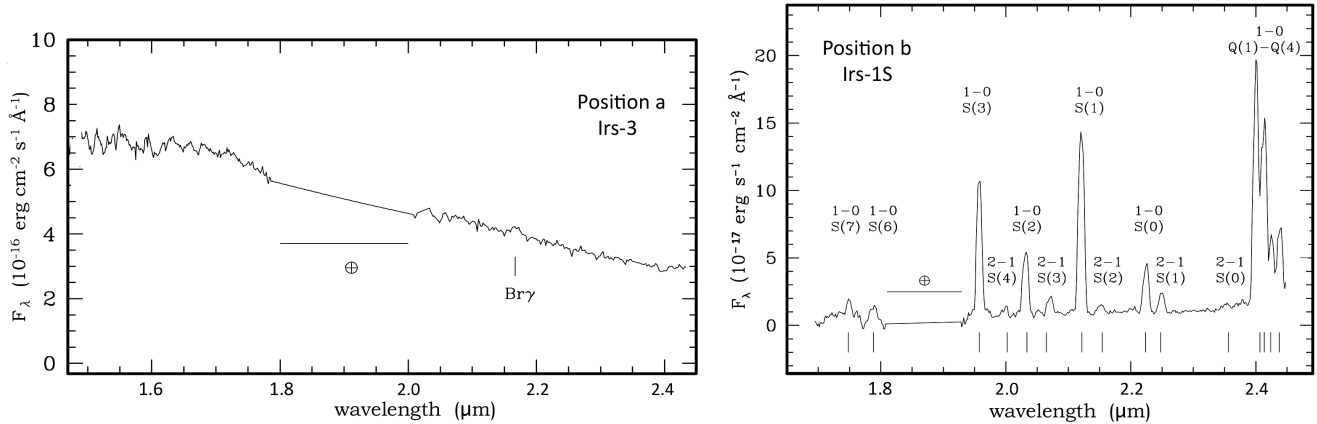}}
\caption{The left panel shows the flux-calibrated spectrum of the central star of the northern HII region (position a in Fig.~5) in the 1.45 to 2.45 $\mu$m 
range. Due to very strong telluric absorption, no data is displayed in the 1.8 to 2.0 $\mu$m wavelength range. The right panel shows the 1.65 to 2.47 
$\mu$m flux-calibrated spectrum of the brightest shocked H$_2$ knot (position b in Fig.~5) and Irs-1S. The identification and measured fluxes of all 
the H$_2$ lines are listed in Table 3.}
\label{fig6}
\end{figure*}
 
 The final calibrated spectra of two bright sources, the star at the centre of the northern HII region (Irs-3) and the second at the spot of peak 
 H$_2$ emission (Irs-1S),  are shown in Fig.~ 6.  The identifications of the emission lines observed in this region with their measured 
 intensities are listed in Table~3.  These results will be further discussed in Sections 3.2 and 3.3. The spectra of several other fainter unresolved 
 sources lying along the slit were  recorded, all showing stellar continuum, with no evident emission lines. These proved to be too noisy for 
 further analyses.

\begin{table}
\caption[] {Identification and measurements of fluxes of the H$_2$ emission lines in the spectrum of the brightest spot.}
\begin{tabular}{lccr}
\hline
Identification & $\lambda$ &Flux \\
&~~($\mu$m)~~& (erg s$^{-1}$ cm$^{-2}$) \\
\hline
H$_2$ 1-0 S(7) & 1.748  & $1.9 \times 10^{-15}$ \\
H$_2$ 1-0 S(6) & 1.788  & $2.3 \times 10^{-15}$ \\
H$_2$ 1-0 S(3) & 1.958  & $1.0 \times 10^{-14}$ \\
H$_2$ 1-0 S(2) & 2.034  & $5.1 \times 10^{-15}$ \\
H$_2$ 1-0 S(1) & 2.122  & $1.5 \times 10^{-14}$ \\
H$_2$ 1-0 S(0) & 2.224  & $3.5 \times 10^{-15}$ \\
H$_2$ 2-1 S(4) & 2.004  & $5.4 \times 10^{-16}$ \\
H$_2$ 2-1 S(3) & 2.074  & $1.3 \times 10^{-15}$ \\
H$_2$ 2-1 S(2) & 2.154  & $6.5 \times 10^{-16}$ \\
H$_2$ 2-1 S(1) & 2.248  & $1.4 \times 10^{-15}$ \\
H$_2$ 1-0 Q(1) & 2.407 & $2.1 \times 10^{-14}$ \\
H$_2$ 1-0 Q(2) & 2.413 & $1.4 \times 10^{-14}$ \\
\hline 
\end{tabular}
\end{table}
% Table 3

\begin{figure}
\resizebox{8.8cm}{!}{\includegraphics[clip=true]{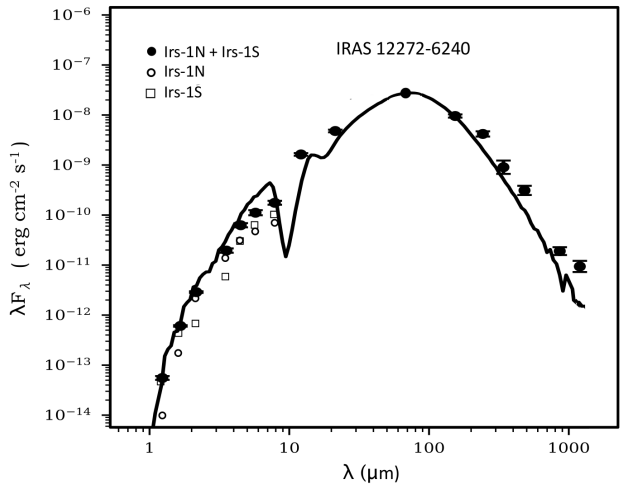}}
\caption{The 1.25 to 1200 $\mu$m SED of IRAS12272-6240 Irs-1 constructed from data in Tables~1 and 2. These were supplemented by $MSX$ 
point source catalogue (Egan et al. 2003) fluxes at 12 and 21 $\mu$m, and from  Csengeri et al. (2013) and Beltr\'an et al. (2006) at 870 and 1200 $\mu$m, respectively.
The best-fit model to the SED of the combined flux densities (Irs~1N + Irs~1S) using Robitaille et al.'s (2007) fitter tool,  is shown by a continuous line.  
The corresponding physical parameters are listed in Table~4. No fit to other models complied with the condition that $\chi^2 - \chi^2_{best}$ per datapoint $< 3$.
Note that for $\lambda > 8 \mu$m, both Irs~1N and Irs~1S are included in their respective apertures.}
\label{fig7}
\end{figure}
 \section{Results and Discussion}

\subsection{Properties of the YSOs in the dense clump and its environment}

The present {\sl Herschel} and {\sl Spitzer} images, displayed in Fig.~1, show the complexity of the IRAS 12272-6240 star formation site. It is dominated 
by a warm (20 K), massive ($1.31 \times 10^4 M_\odot$) dust clump that emits strongly in the far-IR and millimetre wavelengths (Section 2.1). As seen 
in projection in the radio- and mm-continuum, it appears flanked by two round and dusty HII regions (cf. Beltr\'an et al. 2006, S\'anchez-Monge et al. 2013). 
The compact dense clump (DC) houses several CH$_3$OH, OH and H$_2$O masers (Breen \& Ellingsen 2011, S\'anchez-Monge et al. 2013), all indicative of 
massive star formation.

The 1.25 to 1200 $\mu$m SED of IRAS12272-6240 Irs-1 (Irs-1N and Irs-1S combined)  was constructed from data in Tables~1 and 2, supplemented with fluxes 
from the $MSX$ point source catalogue (Egan et al. 2003) at 12 and 21 $\mu$m, and from Csengeri et al. (2013) and Beltr\'an et al. (2006) at 870 and 1200 $\mu$m, 
respectively. Using the in-falling envelope + disk + central star radiation transfer model SED fitter tool by Robitaille \etal (2007), we obtained the best-fit to our 
SED. This is shown by a continuous line in Fig.~7 and the corresponding parameters are listed in Table~4. No fit to other models complied with the condition that 
$\chi^2 - \chi^2_{best}$ per datapoint $< 3$.  Note that for $\lambda > 8 \mu$m, both Irs~1N and Irs~1S are included in their respective beams. Attempts to fit 
plausible SEDs of each individual young stellar component by assuming various fractional contribution to the observed far-IR fluxes proved inconclusive. Their
individual near-IR fluxes are plotted, for illustration, in Fig.~7.

In any case, it is important to stress that one should not over-interpret the results of this widely-used SED model-fitting exercise, regardless of how complete 
and realistic the models themselves may be. The best-fitted parameters constitute only a plausible scenario of the actual physical structure of the young stellar system.
This is especially pertinent to our case of study. Indeed, our high spatial resolution near- and mid-IR images reveal a high degree of complexity at the position 
of the dense clump at the centre of the embedded W cluster (cf. Fig.~3). A detailed look at Fig.~5 evince this.

In spite of the uncertainties involved, it is interesting that the physical parameters of the best-fit to a combined (Irs-1N + Irs-1S) SED are consistent with the 
source being a Class~I massive  YSO with a central O8-9 V star reddened by about $A_V = 15$, not in contradiction the analyses of the near-to-mid-IR
photometric results of the individual components.
Figure 5 presents close-up views of the central area at several wavelengths, with particular attention to the bright Irs-1 protostellar binary system, 
whose northern and southern components are labelled, together with the northern HII region and its central, ionising star, Irs-3.
The colour frame  is coded such that the emission of the ionised hydrogen Br$_\gamma$ line is seen green, the $2.12~\mu$m molecular hydrogen 
H$_2$ line is seen red, and the stellar $K$-band continuum is seen white or blueish. The panel on the right displays, magnified, the comparative contribution 
of the emission in the IRAC bands at 3.6, 4.5 and 5.8 $\mu$m in a small region centred on the binary protostar.  The projected separation between 
Irs-1N and Irs-1S is $2.3\arcsec$ (some 21,000 AU), unresolvable at $\lambda > 8~\mu$m due to the low diffraction-limited spatial resolution of the {\sl MSX},
{\sl WISE} and {\sl Herschel} space telescopes.

Substantial differences are found in the near-IR. Irs-1N displays a large $K_s$-band excess emission that dominates the flux at these 
wavelengths (including a moderate Br$\gamma$ line) over a highly extincted  photosphere. On the other hand, Irs-1S has $J-H$ and $H-K$ colours of a  
less reddened photosphere, with thermal dust emission starting to dominate only at $\lambda > 3~\mu$m. These photometric differences may reflect that 
their individual disc morphologies and inclination angles differ substantially. Another feature that distinguishes between them is the presence of 
extended and bright $2.12~\mu$m molecular hydrogen line emission in the very close vicinity of Irs-1S, absent in Irs-1N.

\begin{figure}
\resizebox{7.5cm}{!}{\includegraphics[clip=true]{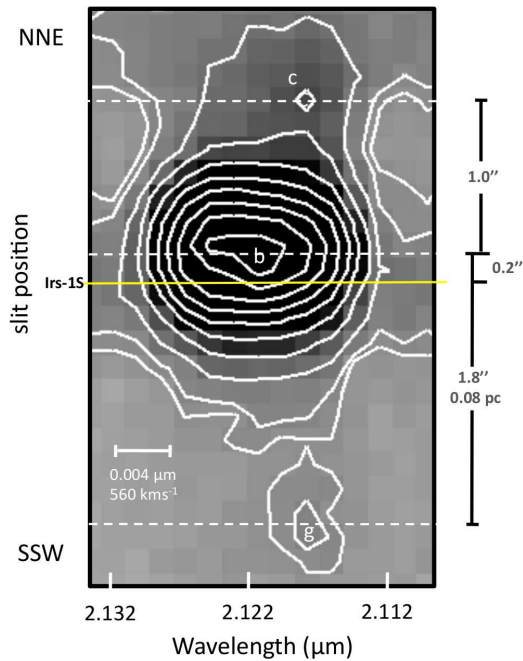}}
\caption{Wavelength versus position greyscale image and linear-scaled contours (for clarity) from a sky-subtracted direct long-slit section of the 
FIRE spectrum centred on the brightest shocked H$_2$ knot (position b in Figs. 3 and 5) and the  $\lambda 2.12~\mu$m emission line. 
The yellow horizontal line, marks the position of the stellar (Irs-1S) continuum, while the white dashed lines mark the positions along 
the slit of the H$_2$ peak emission knots b, c and g. The horizontal white bar represents the nominal spectral resolution of the spectrograph 
with the observational settings. The vertical bars present the values of the separation along the slit between the main emission knots. 
The scale is for a distance of 9.3 kpc.}
\label{fig8}
\end{figure}

\setlength{\tabcolsep}{8pt}

%parameters22april20
 \begin{table}
\caption[] {Physical parameters of IRAS12272-6240 of the best-fit fitting the model.}
\begin{tabular}{lc}
\hline
Parameters&IRAS12272-6240\\
\hline
Stellar mass ($\msol$) & 22.9\\
Stellar temperature (K) & 33500\\
Stellar radius ($\rsol$) & 8.7\\
$A_V$ (magnitudes) & 15.0\\
Envelope accretion rate ($\msol$~yr$^{-1}$)& $8.8 \times 10^{-3}$ \\
Disc mass ($\msol$) & $9.0 \times 10^{-3}$\\
Disc accretion rate ($\msol$~yr$^{-1}$) & $6.1 \times 10^{-7}$\\
Minimum disc radius (AU) & 24 \\
Maximum disc radius (AU) & 61 \\
Inclination angle (degrees) & 32 \\
Fixed distance (kpc) & 9.3\\
Total luminosity ($L_{\odot}$)& $8.5 \times 10^4$\\
\hline 
\end{tabular}
\end{table}
% Table 4

\begin{figure}
\resizebox{8.74cm}{!}{\includegraphics[clip=true]{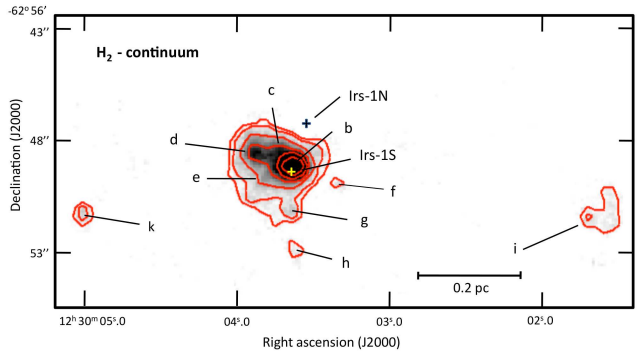}}
\caption{Close-up image of the continuum-subtracted $2.12~\mu$m molecular hydrogen emission line in the vicinity of Irs-1N  (black cross) and 
Irs-1S (yellow cross) YSOs. The contour levels correspond to 5, 9, 18, 36, 61, and 93 times the standard deviation of the background.  The brightest 
nine H$_2$ emission knots (see Section 3.1) are labelled, and their positions are listed in Table 5. The scale is for a distance of 9.3 kpc.}

\label{fig9}
\end{figure}

Our FIRE 1.75 to 2.45 $\mu$m spectrum of Irs-1S (at position b in Fig.~5) shown in the right panel of Fig.~6, is dominated by emission of a 
large number of molecular hydrogen lines, all identified and listed with measured intensities in Table~3. The presence of a faint, very red, stellar continuum, 
and the lack of the Br$\gamma$ line must be noted. Also interesting is the fact that the bright H$_2$ emission is extended and displaced from the underlying 
red star that most probably is responsible for providing the energy for the collisional excitation of the molecule.

In order to study in detail the kinematic structure of this molecular hydrogen emission feature and its likely physical relation to 
the YSO Irs-1S, we present in Fig.~8 a section of the sky-subtracted direct image (wavelength versus position along the slit) of the FIRE long-slit spectrum 
centred on the $2.122~\mu$m line and the position of Irs-1S. To complement this, we present in Fig.~9 the continuum-subtracted $2.12~\mu$m narrow-band 
PANIC image of a small area around Irs-1S. We identified the unambiguous presence of nine compact H$_2$ emission knots (labelled b to k), that 
exhibit  intensity peaks at least 6 standard deviations ($\sigma$) above the background\footnote{The mean random variations of the continuum-subtracted background was 
quantified by calculating the standard deviation, $\sigma$, over the whole frame, after all unresolved (point-like) stars in the field were extracted.}.
These knots of emission are all spatially resolved and have (with the exception of Irs-1S itself) no stellar counterpart (within 0.5 arcsec) 
on the $H$ or $K_s$ images.  Their positions and peak intensities, in terms of the number of $\sigma$s above background, are listed in Table~5. No
other background variations above $3\sigma$ (either positive or negative) are seen on the entire continuum-subtracted image.

\begin{table}
\caption[] {Coordinates of the $2.12~\mu$m H$_2$ line emission knots and of Irs-1N and Irs-1S measured from the continuum-subtracted 2.12$\mu$m image.}
\begin{tabular}{lccc}
\hline
Ident. & R.A. (2000) & Decl.(2000)  & peak intensity\\
  ~   &h\quad m\quad s & $^\circ \quad ^{\prime }\quad ''$ & above noise ($\sigma s)^\dagger$ \\  
 \hline  
b & 12 30 03.63 & -62 56 49.1 &  142 \\  
c & 12 30 03.71 & -62 56 48.3 &   37 \\
d & 12 30 03.88 & -62 56 48.6 &  38  \\
e & 12 30 03.83 & -62 56 49.8 &  20 \\      
f  & 12 30 03.35 & -62 56 49.9 &  7 \\     
g & 12 30 03.65 & -62 56 51.1 &  12 \\    
h & 12 30 03.63 & -62 56 52.9 &   6 \\      
i  & 12 30 01.70 & -62 56 51.4 &   8 \\      
k & 12 30 05.01 & -62 56 51.2 &   10  \\       
Irs-1N & 12 30 03.53 & -62 56 47.2 & -  \\  
Irs-1S & 12 30 03.64 & -62 56 49.4 & -  \\   
\hline 
\multicolumn{4}{l} {NOTE: ~$\dagger ~ \sigma$ is the standard deviation of the background} \\ 
\end{tabular}
\end{table}
%Table 5

Close examination of Fig. 8, that provides kinematic (radial velocity) information, in combination with the morphological details shown in Fig.~9, 
leads to the following conclusions regarding the characteristics of the molecular hydrogen line emission knots in the vicinity of Irs-1S: 
1.~The peak H$_2$  line emission (knot b in Fig.~9) appears round and extended, with a FWHM of $1.05\arcsec$, compared with the FWHM of the
stars in the field, including Irs-1N, of $0.72\arcsec$. 2.~The centroid of this bright knot b is displaced $0.2\arcsec$ to the NNE,
along  the slit, from the stellar continuum position of Irs-1S, a fact that is also evident in the narrow-band images (Fig.~5).  
3.~ Two fainter blobs of emission are found displaced $1.0\arcsec$ to the NNE (knot c) and $1.8\arcsec$ to the SSW (knot g) from the main 
peak but, notably, both are blue-shifted some 600 km~s$^{-1}$ relative to bright knot b. 4.~Other secondary emission peaks  (knots d and e)
stand out from the faint diffuse H$_2$ emission seen extending some $2-3\arcsec$ around Irs-1S.
5.~Other extended knots of emission are located further away. Notably, knots k and j are located  $7\arcsec$ and $12\arcsec$ from
Irs-1S in opposite directions, but not aligned with Irs-1S (cf. Fig.~9). The observed complex distribution of all these knots  cannot be  
explained with the available data. It may be that several, as yet unidentified, fainter YSOs are related to the outflow patterns seen in this region.

\begin{figure}
\center
\resizebox{7.4cm}{!}{\includegraphics[clip=true]{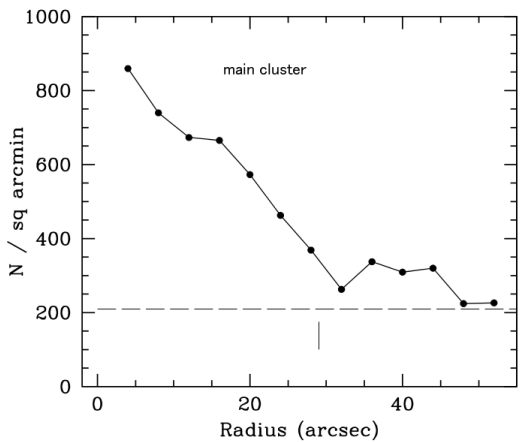}}
\caption{Radial plots of projected number density determined by source counts in concentring rings on the $K_s$ image. The centre was 
determined for the steepest and smoothest function at  $\alpha (2000) = 12^{\rm h}~30^{\rm m}~04\rlap.{^{\rm s}}9$, $\delta (2000) = -62^\circ~56'~50''$. 
The dashed line is the field-star density in this direction determined on the reference sky regions. The small vertical line shows the derived radius of the cluster,
$28\arcsec$.}
\label{fig10}
\end{figure}

\subsection{Properties of the HII regions}

A peculiar feature in our PANIC narrow-band Br$\gamma$ image  is the presence of two  round HII regions (labelled N-HII and S-HII 
in the central and right panels of Fig.~3 and in Fig.~5) emitting brightly in this hydrogen line. N-HII  is centred some $10\arcsec$ to the NNE, and the 
second, S-HII, $30\arcsec$ to the SSW of Irs1. As expected, both HII regions emit strongly also in the IRAC 5.8 and 8 $\mu$m bands, dominated by 
PAHs bands.  Also, there is evidence of considerable amount of warm dust mixed with the ionised gas, as both HII regions are
bright {\sl WISE} sources at 12 (see right panel of Fig.~3) and 22 $\mu$m.

Both HII regions have been measured and characterised by S\'anchez-Monge \etal (2013) based on ATCA 18 and 22.8 GHz radio-continuum observations. 
These authors determined the relevant parameters of the ionised gas. Corrected for the distance of 9.3 kpc, 
N-HII has a radius is 0.26 pc ($5.8\arcsec$), its ionised gas mass  is  $M_{\rm ion} = 2.0\msol$ and 
the number of UV photons $N$(Ly$\alpha$) that it requires is log~N(Ly$\alpha$) = 48.00, equivalent to that produced by an O9 ZAMS (Thompson 1984). 
This is in agreement with Lumsden et al. (2013), who reported a bolometric luminosity of $3.24 \times 10^4 \lsol$, which is that of such O9 star.
Similarly, for S-HII, the radio-derived parameters (S\'anchez-Monge et al. 2013) yield a radius of 1.07 pc ($23.8\arcsec$),  
$M_{\rm ion} = 2.3\msol$ and log~N(Ly$\alpha$) = 48.27, equivalent to an O8 ZAMS. 

\begin{figure*}
\center
\resizebox{18.4cm}{!}{\includegraphics[clip=true]{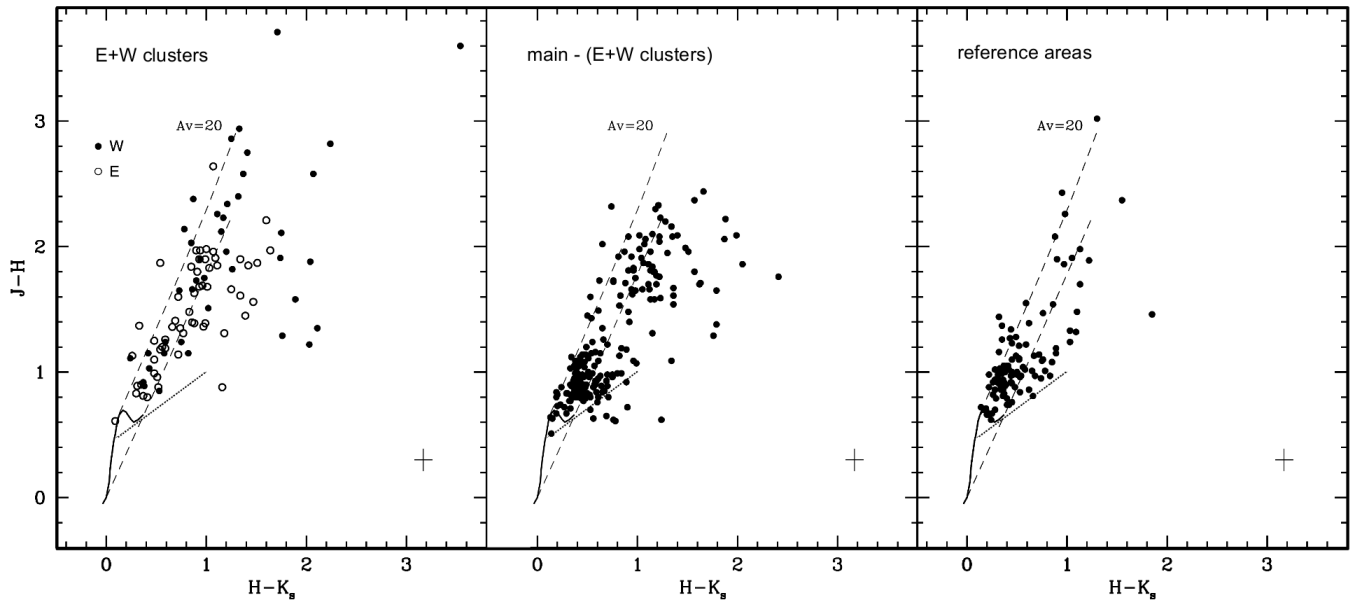}}
\caption{$J-H$ versus $H - K_s$ diagrams of sources measured in $JHK_s$ with uncertainties $\le 0.12$ in each filter. 
The left panel is for the E and W clusters, the central panel is for sources inside the main cluster circle but excluding those in the E and W clusters
(``main - (E+W clusters)''), and the right panel, for sources in the three ``reference'' areas. The different areas covered are defined quantitatively in Table 6, 
and described in Section 3.3. The solid lines mark the loci of the main sequence (Koornneef 1983), the dashed lines delineate the reddening band 
for all main sequence star and giant stars (Rieke and Lebofsky 1985). 
The dotted lines marks the sequence of unreddened Classical T Tauri stars (Meyer, Calvet \& Hillebrand 1997).
The small cross near the lower right corner shows the maximum formal error for each measurement.}
\label{fig11}
\end{figure*}

The boundaries of N-HII are almost circular,  very well defined, suggesting that it is ionisation-bound. The size of the Br$\gamma$ emission measured
on our narrow-band PANIC image is $11.5\arcsec$, confirmed by the measured diameter along the FIRE spectrograph slit of that line emitting region, slightly 
smaller than that measured at 18 GHz. In contrast, the extent of the 2.05 $\mu$m He I line is around $7\arcsec$ in diameter.  The morphology of the 
southern HII region, S-HII, appears quite different in the radio-continuum (S\'anchez-Monge et al. 2013) compared to that seen in the Br$\gamma$ line
(cf. Fig.~3). This suggests the presence of differential dust extinction across the nebula.

We identify Irs-3 as the ionising star of N-HII, located precisely (within $0.2\arcsec$) at the centre of the circular HII region. Its  
$1.45 - 2.45~\mu$m spectrum, shown in the left panel of Fig.~6, appears featureless,  except for a faint Br$\gamma$ line in emission.
This is typical of an O-type star  with a moderated stellar wind. In fact, Its observed colour indices (Table~2) are $J-H = 1.32$ and 
$H-K_s =1.0$ imply the presence of a small near-IR excess emission over that of a reddened O-star photosphere amounting to 
E$_{ff} (H-K_s) \simeq 0.2$ (see left panel of Fig.~11).  This near--IR excess, together with the Br$\gamma$ line, originates in 
the stellar wind plasma. Wind models of massive stars (Kudritzki \& Puls 2000 and references therein) predict this Bremmstrahlung emission
to be negligible at $\lambda \le 1.5 \mu$m. Assuming the measured magnitude $J = 14.72$ of Irs-3 to be representative of the star's photosphere and
the absolute magnitude M$_J = -3.48$ of an O9V star (Martins \& Plez 2006) and Rieke and Lebofsky's (1985) reddening law $A_V = 3.55 A_J$,
at the distance of 9.3 kpc, (true distance modulus (m-M)$_0 = 14.84$) the resulting value of extinction is $A_V = 12.1$. 
Repeating the same exercise for the observed near-IR photometry of the putative ionising star of S-HII, Irs-20,  of a spectral type O8V, 
we obtained  $A_V = 16.4$. Given the uncertainties (in spectral type, absolute magnitude, reddening law), the extinction values are in
concordance with the position of such stars in the two-colour and magnitude-colour diagrams.

\begin{table*}
\caption[] {Coordinates of the centres, and sizes, and star counts ($HK_s$) of the clusters and reference regions.}
\begin{tabular}{lccccccc}
\hline
Name & R.A. (2000) & Decl.(2000) & Radius & Area & Number & Number density & Net number \\
 ~   &h\quad m\quad s & $\circ \quad ^{\prime }\quad ''$  & arcsec & sq. arcmin &  of stars & per sq arcmin & of stars\\  
\hline  
main cluster & 12 30 04.92 & -62 56 50.0 & 28  & 0.684 & 308 & 450 & 166 \\  
E cluster      & 12 30 06.28 & -62 56 45.5 & 9.32 & 0.076 & 58 & 765 & 42 \\
W cluster     & 12 30 03.60 & -62 56 47.0 & 9.32 & 0.076 & 52 & 686 & 36 \\
main-(E+W clusters)  &      -             &         -         &   -   & 0.533 & 198 & 372 & 88 \\
Reference 1 & 12 30 10.55 & -62 56 17.0 & 15  & 0.196 & 32  & 163 & - \\      
Reference 2 & 12 29 57.60 & -62 56 36.0 & 15  & 0.196 &  44  & 224 & - \\     
Reference 3 & 12 30 01.50 & -62 56 34.0 & 15  & 0.196 &  46  & 235 & - \\
Reference 1+2+3 &       -     &     -             &   -    & 0.589 & 122  & 207 &  0 \\
\hline 
\end{tabular}
\end{table*}
% Table 6

\subsection{Properties of the main embedded cluster and sub-clusters}

By means of star-counts on our full catalogue of $HK$-detected sources and assuming a circular projected morphology (see Tapia et al. 2014 for a 
description of the method), we determined the centre and extension of the ``main'' star cluster.  Fig.~10 shows the radial projected source density plot 
of all $HK_s$-measured sources with a well-defined number density peak, which we identify as the cluster centre at 
(J2000)  $\alpha  = 12^{\rm h}~30^{\rm m}~04\rlap.{^{\rm s}}9$, $\delta = -62^\circ~56'~50''$. The star number density decreases   
outwards monotonically until, at a radius of  $28\arcsec$, reaches (within $2\sigma$) the value of the mean projected density of 
207 sources per square arcmin, which is also that mean density in the observed field outside this circle. 
We, thus, adopt the radial extension of the ``main'' cluster to be $28\arcsec$, 
corresponding to a diameter of 2.6 pc at  $d = 9.3$ kpc. This result confirms the initial suggestions by Mercer et al. (2005) of the presence of an 
IR embedded cluster associated with IRAS12272-6240. We also estimated the centres and sizes of two sub-clusters (E and W) that correspond 
to clearly discernible local bright overdensities, each with a diameter of $18.6\arcsec$, or 0.84 pc, with their centres separated by $16.3\arcsec$.
These limits are illustrated on the PANIC and the IRAC mosaics displayed in Fig.~3 (left and right panels). 

\begin{figure*}
\center
\resizebox{15cm}{!}{\includegraphics[clip=true]{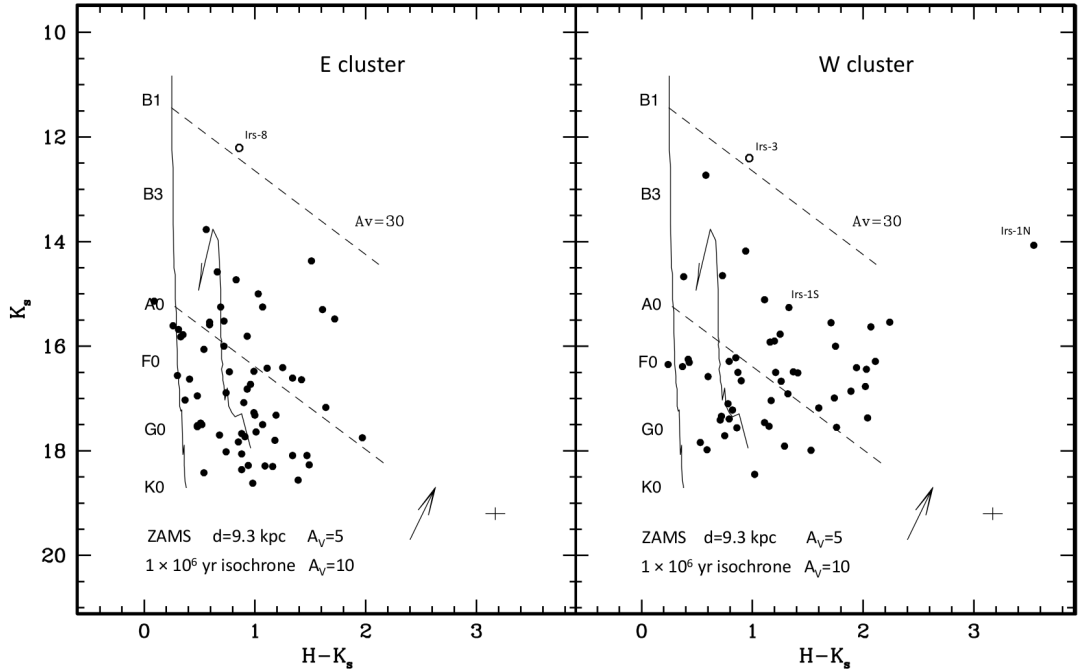}}
\caption{$K_s$ versus $H-K_s$ diagram of sources measured with $H$ and $K_s$ with uncertainties $\le 0.12$ in each filter, including those
with $J$ magnitudes fainter than our detection limit. The left panel is for sources in the E cluster and the right panel is for sources in the W cluster. 
The different areas covered are defined quantitatively in Table 6, and described in Section 3.3.
The zero-age main sequence (ZAMS) for $d = 9.0$~kpc and $A_V = 5$ is represented by almost-vertical solid lines. The curved line is the 1Myr isochrone from 
Siess et al. (2000) reddened by $A_V = 10$ at same distance from the Sun. The dashed lines are the reddening vectors of length 
$A_V=30$ (Rieke and Lebofsky 1985). The small cross near the lower right corner shows the maximum formal error for each measurement. The arrows represent the 
average slope of the near-IR emission excess caused by discs around YSOs, as determined by L\'opez-Chico \& Salas (2007). }
\label{fig12}
\end{figure*}

\begin{figure*}
\center
\resizebox{15cm}{!}{\includegraphics[clip=true]{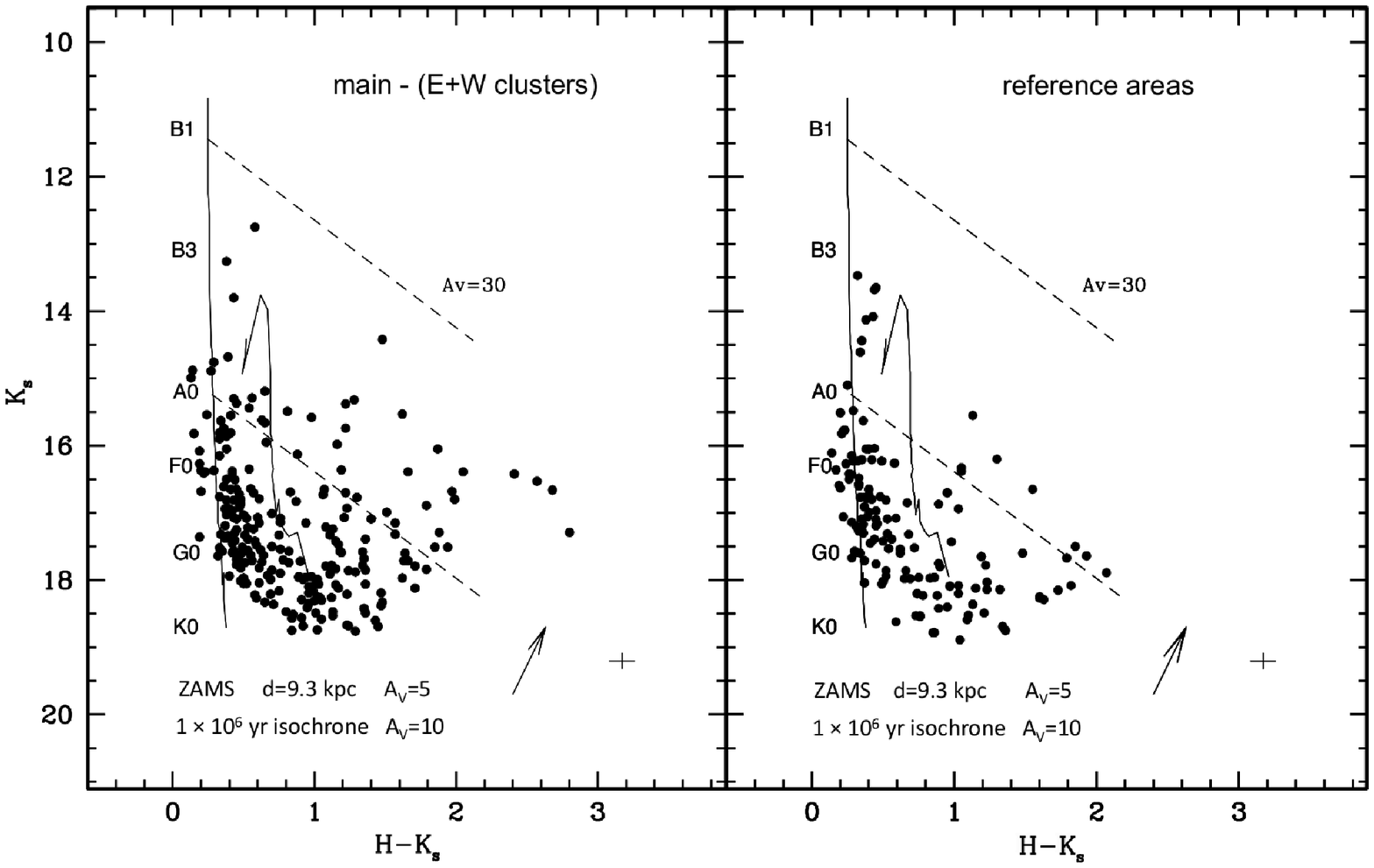}}
\caption{Same as Figure12 but for the ``main-(E+W)'' sources (left panel) and  those in the three reference areas (right panel). The different areas 
covered are defined quantitatively in Table 6  and described in Section 3.3.}
\label{fig13}
\end{figure*}

The present photometric data indicates a considerable complexity in terms of a variety of the amount of dust extinction and of stellar population ages 
from cluster to cluster. In this subsection we present evidence that all of them are part of a single giant young complex that has 
suffered several episodes of star formation. These started a few million years ago and the latest is still active in the dense clump. 
Clearly, we cannot determine what is the line-of sight separation between sub-clusters, but we are able to provide evidence that 
all these structures are roughly at the same distance from the Sun ($\sim 9.3$ kpc) and, thus, form a single giant star formation complex.

In order to perform a comparative analysis of the richness and stellar populations of the embedded clusters, we divided our sample 
into three zones of interest:  The E cluster, the W cluster, and the remainder population within the ``main'' cluster circle (of $r = 28\arcsec$), 
that is, the ``main - (E+W)'' cluster. For comparison purposes, we also defined three ``reference'' circular regions, each of radius $15\arcsec$, 
scattered outside the main cluster.  Table~6 lists the centres and radii of each of these circular regions as well as the resulting star-counts, 
including (in the last column) the net number of stars in each cluster. The latter was calculated by subtracting the expected number of field 
stars, i.e. those counted in the three reference regions normalised by the area of each cluster.

The $J - H$ versus $H - K_s$ diagrams of the stars bright enough to have reliable $J$-band photometry in the four areas defined above are
presented in Fig.~11. The corresponding $K_s$ versus $H - K_s$ diagrams for our whole sample are shown in 
Figs.~12 and 13.  

Consider first  the $J - H$ versus $H - K_s$ and the $K_S$ versus $H-K_s$ diagrams of the (combined) three reference regions shown in the 
right panels of Figs.~11 and 13. This sample is expected to provide information on the characteristics of the field population in the observed direction 
($l = 300.50, b=-0.17$).  It is well-established (eg. Jones at al. 1981) that in a brightness-limited near-IR survey on the Galactic plane, the detected population
along the line of sight will be dominated by late-type stars, as dictated by the $JHK$ field luminosity function. 
In fact, in the $J - H$ versus $H - K_s$ diagram of the reference areas, we find that 74\% of the stars occupy the locus defined by late-type 
photospheres reddened by up to $A_V = 5$, presumably of the foreground, while a handful more (7\%) lie along the same reddening vector, with
$5 <  A_V < 20$, belonging to the background. A further 9\% appear to the right of the reddening vector for early-type stars, implying moderate near-IR excess. 
When we consider their rather faint $K_s$ magnitudes, they appear to be background late type stars with some chromospheric activity.
In both the two-colour and the colour-magnitude diagrams, the observed distributions are clearly different from those seen in the corresponding  diagrams 
for the cluster areas (Figs.~11, 12 and 13). For this reason, we feel confident to assume that, statistically, the reference regions represent the field star population.

\subsubsection{The W cluster}

This small cluster (0.85 pc in diameter) contains at its centre the compact dense clump with the Class-I YSOs described in Section 3.1.  
Perhaps not surprisingly, this region contains the reddest near-IR objects that we measured in this survey, including the Class~I YSOs 
Irs-1N and Irs-1S. In fact, we found that about 80\% of the net (field-subtracted) population of this cluster has $H-K_s > 1.2$. 
(right panel of Fig.~12). This effect is unlikely to be caused by enhancement of dust extinction towards those stars, but rather to 
the presence of discs emitting strongly at $\lambda \ge 2~\mic$ in a significant fraction of the W cluster young members. The support for 
this statement comes from the fact that more than 50\% of the W cluster stars occupy a zone in the $J-H$ versus $H-K_s$ diagram (Fig.~11)
that indicates $K_s$ excess emission of at least 0.8 magnitudes over that expected for reddened photospheres. 
In fact, the fraction of YSO with discs in this cluster may be even larger, once contamination
from the older ``main'' and E clusters is taken into account, as will be discussed in Section 3.3.2. Such a large fraction of YSO with discs 
implies an age well below one million years for the W cluster. Evidently, the Class~I objects Irs-1N and Irs-1S are the brightest and 
youngest protostars at its nucleus.

\subsubsection{The E and ``main'' cluster}

The quite large extension of the ``main'' cluster, determined by means of radial star counts (Fig.~10), implies that the amount
of field stars contamination is severe, especially in the faint end of the observed population. On the other hand, it is clear that their brightest
stars (at $2.2~\mic$) are concentrated in the area that we have defined as the E cluster. The magnitude-colour diagram (left panel of 
Fig.~12) of the latter shows, once the field stars are ignored, a heavy concentration of stars around a $10^6$-year isochrone that is 
reddened by $A_V \simeq 10$ (i.e.  an excess of 5 magnitudes over the foreground extinction), having $0.6 < H-K_s < 1.0$. 
Furthermore, a few (around 10) stars having colour indices in that range lie within the W cluster boundaries  but could very 
well be contaminant members of the E cluster. These would  increase to nearly 60\% the fraction of stars of the latter lying very close 
to the 1 Myr isochrone (cf. Fig.~12).   We, thus, naturally conclude that  the mean age of the embedded W cluster is about  $10^{6}$ years. 
Fitting into this picture, the central stars of the N- and S-HII regions (Irs-3 and Irs-20) would be the most massive members of the W cluster. 
Note, on the other hand, that ignoring the contamination of field stars,  the colour-magnitude diagram of the more extended, 
``main - (E+W) cluster'' (Fig.~13) does resemble a mixture of the combined E and W cluster populations, but with a much lower 
number density of stars. We envisage this population to constitute a fainter extension (or halo) of the E and W denser clusters. 
  
\section{Conclusions}

In this work, we present new sub-arcsec broad- and narrow-band near-IR imaging  of a 120 x 120 square arcsec area centred 
on the massive star-forming region IRAS 12272-6240. We supplemented these data with HI-GAL/{\sl Herschel} far-IR images 
combined with archive IRAC/{\sl Spitzer} and {\sl WISE} mid-IR observations.  We also present low-resolution near-IR spectroscopy
of a couple of characteristic sources in the complex. The analyses of these observations yielded the following main results:

1.  IRAS 12272-6240, located at a distance of 9.3 kpc from the Sun, consists of a compact dense clump that houses two Class~I YSOs, 
Irs-1N and Irs-1S, probably forming a binary system, with several close CH$_3$OH, OH and H$_2$O masers associated. We found the 
presence of a series of compact molecular hydrogen emission knots in its close vicinity, further indicating the existence of strong molecular 
outflows. Robitaille \etal's (2007) model best-fitted to its 1 to 1200 $\mu$m SED is consistent with a $23 \msol$ central star with 
an effective temperature of 33,500 K and a $10^{-2} \msol$ disc at an inclination angle of $32\degr$. The dust envelope has 
a mass of $1.3 \times 10^4 \msol$ and a mean temperature of 20 K. Its total luminosity is $8.5 \times 10^4 \lsol$. 
 
2. Based on our near-IR colour-magnitude and the fractional distribution of stars with evidence of having pre-planetary discs,
we conclude that in the large star-forming complex IRAS 12272-6240, we see two embedded clusters differing in age, 
spatial distribution and physical characteristics. The older and more extended of these, 
the E cluster, has an age of about one million years, and is reddened by $A_V \simeq 10$. It contains more than 50 stars in its 
nucleus and a halo of some 80 fainter stars extending to a radius of about 1.3 pc (the ``main'' cluster).  We distinguished another, more 
compact W cluster, of which we measured more than 35 members,  that appears more deeply embedded and has a younger stellar population, 
of which a large fraction ($> 50\%$) exhibits considerable $K$-band excesses that evince the presence of discs. Based on this, we estimate the
W cluster to be significantly younger than $10^6$ years. This cluster includes at its centre the massive YSOs Irs-1N and Irs-1S. 

3. Two round, dusty radio/IR HII regions, named S-HII and N-HII, flank IRAS 12272-6240.  We identified in our IR images their central stars and, 
based on their IR photometry and distance, we derived their spectral types. S-HII has an O8V central star with $A_V \simeq 16$, while the 
star ionising N-HII has a spectral type O9V, and is reddened by  $A_V \simeq 13$. These values correspond precisely to the ionising stars 
required to produce their observed radio-continuum fluxes. At radio wavelengths, S-HII has a diameter of 2.0 pc, and looks asymmetrical in
the near-IR. In contrast, the appearance of N-HII is similar in the radio-continuum and in the Br$\gamma$ line, with a diameter of 0.5pc.
The stellar photometry and derived distance moduli imply that both are members of the E and ``main'' cluster. 

4. The present results confirm that that all the elements described above form a single giant young complex located 9.3 kpc from the Sun, and 
where massive star formation processes started some 1 million years ago and is still active at its nucleus. Further high-resolution observations, 
such as in the X-ray regime, are needed to wide the search for the young low-mass population in the region IRAS 12272-6240 in order to test
for possible mass and age segregation, as it has been reported in other massive star-formation sites, such as DR 21 (Rivilla et al. 2014).

\section{Acknowledgements}

The authors acknowledge an anonymous referee for her/his comments and suggestions that led to a considerable improvement in the clarity and 
content of this paper. MT acknowledges support for this work through  PAPIIT-UNAM grant IN-107519. This paper makes use of archival data obtained with
the {\sl Spitzer Space Telescope}, which is operated by the Jet Propulsion Laboratory, California Institute of Technology (CIT)
under National Aeronautics and Space Administration (NASA) contract 1407. It also makes use of data products from the {\sl Wide-field Infrared Survey 
Explorer (WISE)}, which is a joint project of the University of California, Los Angeles, and the Jet Propulsion Laboratory/California Institute of Technology, 
funded by the National Aeronautics and Space Administration. This research has made use of the NASA/IPAC Infrared Science Archive (IRSA), which is 
also funded by the National Aeronautics and Space Administration and operated by the California Institute of Technology.

\section{Data availability}

The data underlying this article are available in the article and, electronically, through CDS at http://cdsarc.u-strasbg.fr/viz-bin/qcat?J/MNRAS/.

\end{document}